\documentclass[twocolumn, superscriptaddress, preview, amsmath, amssymb, aps, prl, letterdraft, notitlepage]{revtex4-1}
\usepackage[utf8]{inputenc}
\usepackage{natbib}
\usepackage{graphicx}
\usepackage{mathtools}
\usepackage{xcolor}
\usepackage{comment}

\newcommand{\bra}[1]{\langle#1|}
\newcommand{\ket}[1]{|#1\rangle}

\newcommand{\be}{\begin{equation}}
\newcommand{\ee}{\end{equation}}

\newcommand{\avg}[1]{\left \langle #1\right \rangle}

\newcommand{\timeder}{\frac{\partial}{\partial t}}

\begin{document}
\title{Variational Quantum Simulations of Finite-Temperature Dynamical Properties via Thermofield Dynamics}

\author{Chee Kong Lee}
\email{cheekonglee@tencent.com}
\affiliation{Tencent America, Palo Alto, CA 94306, United States}
\author{Shi-Xin Zhang}
\affiliation{Tencent Quantum Lab, Shenzhen, Guangdong 518057, China}
\author{Chang Yu Hsieh}
\affiliation{Tencent Quantum Lab, Shenzhen, Guangdong 518057, China}
\author{Shengyu Zhang} 
\affiliation{Tencent Quantum Lab, Shenzhen, Guangdong 518057, China}
\author{Liang Shi}
\email{lshi4@ucmerced.edu}
\affiliation{Chemistry and Biochemistry, University of California, Merced, California 95343, United States}

\begin{abstract}
The recent advancement of quantum computer hardware offers the potential to simulate quantum many-body systems beyond the capability of its classical counterparts. 
However, most current works focus on simulating the ground-state properties or pure-state quantum dynamics of quantum systems.  Finite-temperature physics are much less explored, and existing literature mainly consider equilibrium properties. 
Here we present a variational quantum simulation protocol based on the thermofield dynamics formalism to simulate the dynamical and non-equilibrium finite-temperature properties of quantum systems with near-term quantum computers.
Compared to previous approaches in computing the equilibrium dynamical properties, our approach does not require the computationally expensive initial state sampling. 
More importantly, our approach is capable of simulating non-equilibrium phenomena which have not been previously explored with quantum computers. 
Numerical simulations of molecular absorption spectrum and spin quenched dynamics affirm the capability of our method. 
\end{abstract}

\maketitle

Despite many decades of advancements, simulations of quantum many-body systems remain one of the most challenging problems in physics. The rapid improvement in quantum computing hardware offers the potential to tackle such challenges beyond the capability of any classical computers~\cite{Arute2019, Wu2021}. 
While there has been much progress in near-term quantum simulation of ground-state properties and pure-state quantum dynamics of many-body systems~\cite{Cerezo2021, Bharti2022}, finite-temperature physics has been much less explored. 
Earlier works on digital quantum simulation of finite-temperature physics typically require additional ancilla qubits and rely on the existence of fault-tolerant quantum computers due to the long circuit depth~\cite{Poulin2009, Riera2012, Yung2012, Temme2011}, thus they are not suitable for near-term noisy intermediate-scale quantum (NISQ) devices in which quantum resources are limited. 

In recent years, there have been several near-term quantum algorithms exploring finite-temperature physics, such as algorithms based on the quantum imaginary-time evolution (QITE)~\cite{Motta2020}, thermofield double states~\cite{Wu2019}, and neural networks~\cite{Chowdhury2020, Liu2021, Zoufal2021}.
The majority of these works focus only on the static equilibrium properties, despite the ubiquity of dynamical and non-equilibrium phenomena in physics and chemistry. 
Only recently has the QITE algorithm been extended to compute the dynamical (still equilibrium) properties~\cite{Sun2021}. However, the algorithm requires sampling of initial states in computing observables, and the cost of such sampling scales unfavorably with system sizes. Specifically, the sampling cost of the exact full trace evaluation scales exponentially with systems sizes whereas the scaling of the approximate stochastic trace evaluation remains unclear. 
Furthermore, to the best of our knowledge, quantum algorithms for simulating non-equilibrium finite-temperature physics has not been explored.

Here we adopt an extension of the thermofield double state formalism to perform quantum simulations of equilibrium and non-equilibrium dynamical properties at finite temperatures. 
The thermofield double states formalism has been used to study various finite-temperature phenomena, e.g. black holes~\cite{Maldacena2013} and finite-temperature quantum chemistry~\cite{Harsha2019}.  
Recently it has been adopted for quantum simulation of equilibrium systems via a variational hybrid quantum-classical algorithm~\cite{Wu2019}. The hybrid algorithm has subsequently been experimentally demonstrated using the ion-trap~\cite{Zhu2020} and superconductor quantum computers~\cite{Sagastizabal2021}. 
Similar to the equilibrium thermofield double states, finite-temperature quantum systems within the thermofield dynamics (TFD) framework are represented by pure quantum states entangled between the physical system of interest and a fictitious system, the time-dependent density matrix of the system of interest can then be obtained by tracing over the fictitious degrees of freedom.
The equation of motion for the TFD wavefunction is governed by the Hamiltonian of the combined systems. The TFD formalism has recently adopted to simulate exciton dynamics~\cite{Borrelli2017}, non-adiabatic quantum dynamics~\cite{fischer2021a}, and spectroscopy~\cite{gelin2021b,chen2021b,Begusic2021,polley2022}. 

We combine the TFD framework with a hybrid variational quantum-classical algorithm to enable its implementation on NISQ devices in a straight-forward manner.
As opposed to the QITE formalism, our approach does not incur additional initial state sampling cost albeit it requires additional qubits in representing the fictitious systems. 
To demonstrate the capability of our approach, we report the numerical simulations of the quenched dynamics of a spin system and the absorption spectroscopy of a molecular system, and the excellent agreement between the results from our method and exact calculations demonstrates the accuracy of our approach.  

\textit{Thermofield dynamics:} We consider a general time-dependent Hamiltonian, $H(t)$, which governs the time evolution of the density matrix, $\rho(t)$, of a physical system according to the Liouville–von Neumann equation:
\be
i \timeder \rho(t) = [H(t), \rho(t)],
\label{eq:lvn}
\ee
where $\hbar=1$ is used throughout the paper.
The TFD formulation of quantum mechanics extends the original Hilbert space spanned by an orthonormal complete basis, $\{\ket{\alpha}\}$, to a double space, and constructs a purification of the density matrix by introducing a time-dependent thermal quantum state in the double space,\cite{suzuki1985,suzuki1986,suzuki1991}
\be
\ket{\Psi(t)} = d^{1/2} \rho(t)^{1/2} \ket{I},
\label{eq:def}
\ee
where $\ket{I} = \sum_{\alpha=1}^d \ket{\alpha}\ket{\tilde{\alpha}}/\sqrt{d}$, $d$ is the dimension of the Hilbert space of the physical system, and $\ket{\tilde{\alpha}}$ denotes the basis state for the fictitious system which is identical to the physical system. With such purification, the evolution of $\rho(t)$ in the original space is equivalent to the evolution of $\ket{\Psi(t)}$ in the double space following\cite{suzuki1985,suzuki1986,suzuki1991,takahashi1996}
\be
i \timeder \ket{\Psi(t)} = \hat{H}(t) \ket{\Psi(t)},
\label{eq:eom_psi}
\ee
where $\hat{H}(t) = H(t)  - \tilde{H}(t)$, and $\tilde{H}(t)$ is the Hamiltonian of the fictitious system, which is identical to $H$. In deriving Eq. (\ref{eq:eom_psi}), one realizes that the operators in the physical and fictitious system commute, and uses the equation of motion for $\rho(t)^{1/2}$:~\cite{ezawa1990}
\be
i \timeder \rho(t)^{1/2} = [H(t), \rho(t)^{1/2}].
\ee
The definition of $\hat{H}(t)$ formally corresponds to a forward time evolution of the physical system and a backward time evolution of the fictitious system.
The density matrix of the physical system can be recovered by taking the trace over the fictitious system, $\rho(t) = \textrm{Tr}_f \ket{\Psi(t)}\bra{\Psi(t)}$.
For an arbitrary operator, $A$, of the physical system, its statistical average is now given by its expectation value with respect to the time-dependent thermal quantum state,
\be
\avg{A}_t = \textrm{Tr} \{ \rho(t) A \} = \bra{\Psi(t)}  A  \ket{\Psi(t)},
\ee
where the subscript $t$ in $\avg{A}_t$ indicates the time dependence of the non-equilibrium statistical average.

If the system is at thermal equilibrium, $\rho(t) = \rho_{\textrm{eq}} = e^{-\beta H}/Z$, where $Z(\beta) =\textrm{Tr} \{e^{-\beta H}\}$, and $\beta = 1/k_B T$, and $\ket{\Psi(t)}$ recovers the so-called thermo-field double state (also known as thermal vacuum state),
\be
\ket{O(\beta)} = d^{1/2} Z(\beta)^{-1/2} e^{-\beta H/2} \ket{I}.
\ee
The thermal equilibrium average of an arbitrary operator, $A$, of the physical system is then given by
\be
\avg{A} = \textrm{Tr} \{ \rho_{\textrm{eq}} A \} = \bra{O(\beta)}  A  \ket{O(\beta)}.
\ee

\textit{Variational Quantum Algorithm:} The TFD theory maps the evolution of the density matrix to that of a wave function in the double Hilbert space, which is more amenable to implementation on a quantum computer. Here, we adopt a hybrid variational quantum algorithm (VQA) based on McLachlan's variational principle~\cite{Li2017, Cerezo2021,yuan2019,endo2020a,endo2021} to initialize and evolve the TFD state $\ket{\Psi(t)}$, which are suitable for NISQ devices. For real-time evolution, a trial quantum state $\ket{\psi(\vec{\theta}(t))}$ with parameters $\vec{\theta}(t)$ evolves to minimize the quantity $||(i \partial_t - \hat{H}(t))\ket{\psi(\vec{\theta}(t))}||$, and the resulting equation of motion for $\vec{\theta}(t)$ is~\cite{yuan2019,endo2020a,endo2021} 
\begin{equation} \label{eq:eom}
    \sum_l \textrm{Re} \{M_{kl}\} \dot{\theta}_l = - \textrm{Im}\{V_k\},
\end{equation}
where the matrix elements of $M$ and $V$ are
\begin{eqnarray} \label{eq:elements}
 M_{kl} =    \left\langle \frac{\partial \psi(\vec{\theta})}{\partial \theta_k} \right\vert \left. \frac{\partial \psi(\vec{\theta})}{\partial \theta_l}
\right\rangle, \\
 V_{k} =    \left\langle  \psi(\vec{\theta}) \right\vert  \hat{H} \left\vert \frac{\partial \psi(\vec{\theta})} {\partial \theta_k }
\right\rangle.  
\end{eqnarray}
$M_{kl}$ and $V_{k}$ can be measured on quantum computers using the standard Hadamard test, and the quantum circuits for them are relatively shallow as shown in SM, thereby suitable for NISQ devices. The updating of the parameters $\vec{\theta}$ will be performed on classical computers.

In many cases, the initial TFD state is the thermofield double state, $\ket{\Psi(0)} = \ket{O(\beta)}$, which can be prepared using an imaginary-time VQA based on McLachlan's variational principle for the Wick-rotated Schr\"{o}dinger equation.~\cite{yuan2019,endo2020a,endo2021}  The initial state for the imaginary-time evolution is the maximally entangled state, $\ket{I}$, and its imaginary-time evolution is approximated by a parameterized state $\ket{\psi(\vec{\theta}(\tau))}$, where $\tau$ denotes the imaginary time. To obtain $\ket{O(\beta)}=d^{1/2} Z(\beta)^{-1/2} e^{-\beta H/2} \ket{I}$, the imaginary-time evolution is performed up to the imaginary time $\tau = \beta/2$ according to the following updating equation for $\vec{\theta}$,~\cite{yuan2019,endo2020a,endo2021} 
\begin{equation} \label{eq:im_eom}
    \sum_l \textrm{Re} \{M_{kl}\} \dot{\theta}_l = - \textrm{Re}\{V_k\}.
\end{equation}
Note that $\textrm{Re}\{V_k\}$ is related to the gradient of the energy of the physical system by
\be
\textrm{Re}\{V_k\} = \textrm{Re} \left\langle  \psi(\vec{\theta}) \right\vert  H \left\vert \frac{\partial \psi(\vec{\theta})} {\partial \theta_k }  \right\rangle = 
\frac{1}{2} \frac{\partial}{\partial \theta_k} \bra{\psi(\vec{\theta})} H \ket{\psi(\vec{\theta})}. 
\ee

In this work we consider wavefunction ansatz of the form $\ket{\psi (\vec{\theta})} = U(\vec{\theta}) \ket{\psi_0} = \prod_{k=1}^{N_\theta} \mbox{e}^{i \theta_k R_k} \ket{\psi_0}$
where $\ket{\psi_0} = \ket{I}$ is the initial wave function, $R_k$ is some Pauli string, and $N_\theta$ represents the number of variational parameters. The circuit depth for preparing such ansatz is therefore proportional to the number of variational parameters.  
The measurement cost is dominated by the matrix $M$ and scales with $N_\theta^2$. It is worth noting that the methodology developed in our work is general and does not depend on the specific form of wavefunction ansatz.
More details about the implementation of the real and imaginary-time VQA can be found in the SM.

\textit{Quenched Spin Dynamics:} To demonstrate the capability of the thermofield VQA in simulating finite-temperature non-equilibrium phenomena, we first consider the time evolution of a one-dimensional transverse-field Ising (TFI) model induced by quantum quench. The TFI Hamiltonian is 
\begin{eqnarray}
    H_{\textrm{TFI}} &=& - h \sum_{i} \sigma_i^{x} - \sum_{<ij>} \sigma_i^{z} \sigma_j^{z},
\end{eqnarray}
where $h$ denotes the strength of the transverse field. 
The TFI system is initially prepared in the thermal state of an initial transverse field $h_i$, $\rho(0) = e^{-\beta H_{\textrm{TFI}}}/Z$.
In the TFD framework, the initial thermal state can be represented by the double state $\ket{O(\beta)}$, which in turn can be approximated in a quantum circuit with a wavefunction ansatz. 
Specifically, the variational parameters of the initial thermal double state are obtained using the imaginary-time algorithm (see Eq.~(\ref{eq:im_eom})).

\begin{figure}[ht!]
    \centering
  \includegraphics[width=0.50\textwidth]{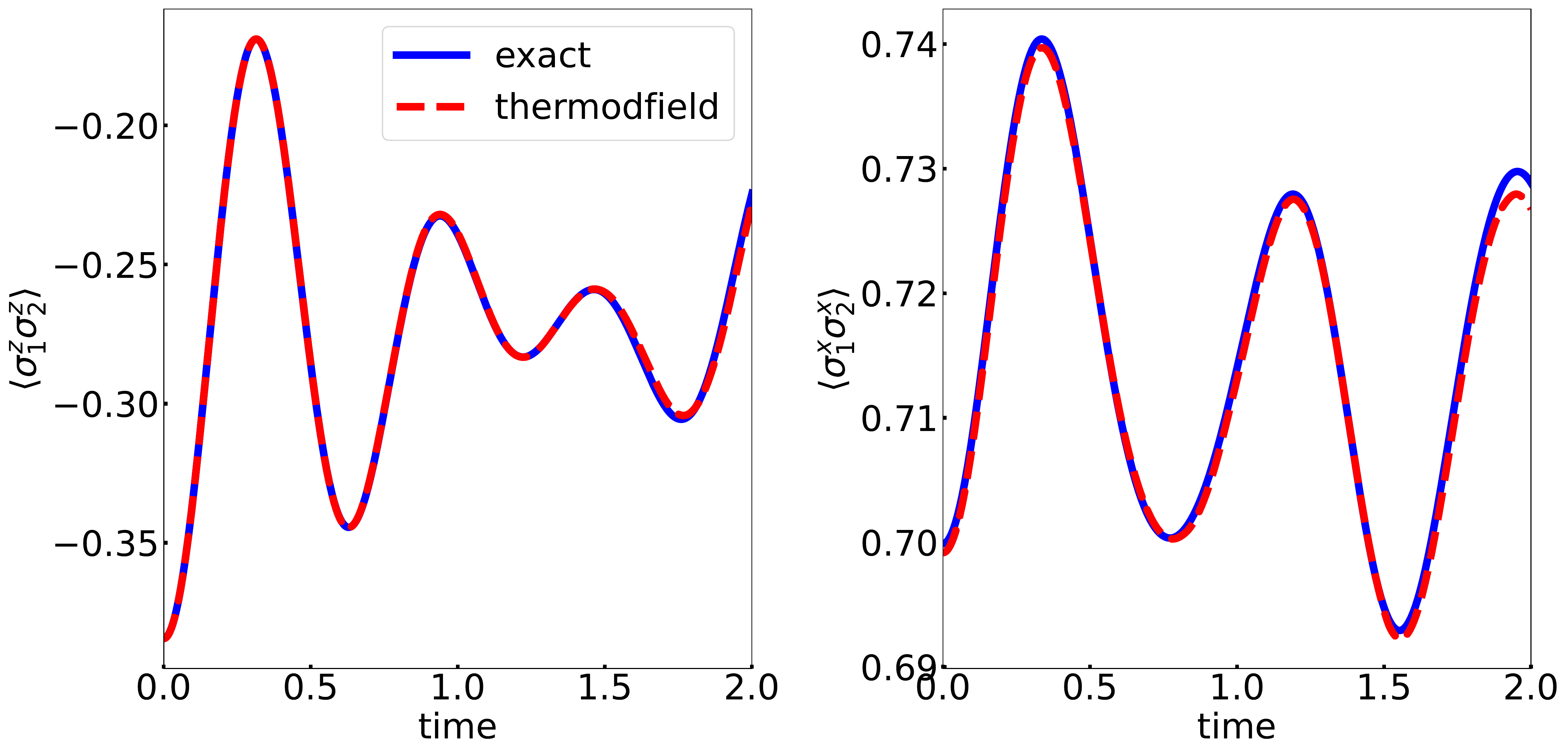}
    \caption{Time evolutions of the magnetization (left) and transverse polarization (right) correlations of a four-spin TFI model system, induced by a quantum quench from $h_i=1.5$ to $h_f=2.5$ at $t=0$. The inverse temperature is set to be $\beta=0.5$. Results from the TFD-based variational quantum algorithm (blue dashed lines) are compared with numerically exact calculations (red solid lines).}
    \label{fig:TFI}
\end{figure}

At $t=0$, we introduce an instantaneous change to the transverse field, $h_f \neq h_i$, and let the system evolve under the new Hamiltonian. In Fig. ~\ref{fig:TFI}, we consider a TFI system of four physical spins, and the parameters used are $h_i = 1.5$, $h_f = 2.5$ and $\beta = 0.5$. 
We compare the results from the TFD-based VQA with those from exact numerical calculations by studying the evolution of magnetization correlation, $\langle\sigma_i^{z} \sigma_j^{z} \rangle$, and transverse polarization correlation, $\langle\sigma_i^{x} \sigma_j^{x} \rangle$. The exact numerical results are obtained by numerically evolving the Liouville-von Neumann equation (Eq.~(\ref{eq:lvn})). The good agreement with exact results in Fig. ~\ref{fig:TFI} confirms the accuracy of the variational algorithm in capturing non-equilibrium finite-temperature quantum dynamics.

\textit{Correlation Function and Spectroscopy:} Next we use the TFD formalism to evaluate finite-temperature equilibrium time correlation function:
\begin{eqnarray}
    C(t) = \textrm{Tr}\{\rho_{\textrm{eq}} A(t) B \},
\end{eqnarray}
where $\rho_{\textrm{eq}} = \mbox{e}^{-\beta H}/Z$, $A(t) = U^\dagger(t) A U(t)$, and $U(t) = \mbox{e}^{- i H t}$. In the context of TFD, the correlation function can be written as 
\begin{eqnarray}
    C(t) = \bra{O(\beta)} \hat A(t) B \ket{O(\beta)}.
\end{eqnarray}
where $ \hat A(t) = \hat{U}^\dagger (t) A \hat{U} (t)$ and $\hat{U} (t) = \mbox{e}^{- i \hat H t}$.

Here we first prepare the thermal state variationally using the imaginary-time VQA, i.e. $\ket{O(\beta)} \approx \ket{O(\beta(\vec{\theta}))} = U(\vec{\theta}) \ket{I}$. We then propagate two sets of wavefunctions using the real-time VQA (see Eq.~(\ref{eq:eom})): 
\begin{eqnarray}
    \ket{\psi_1(t)} &=& \hat U(t) B \ket{O(\beta)} \approx U_1(\vec{\theta}_1) B \ket{O(\beta(\vec{\theta}))}, \\
   \ket{\psi_2(t)} &=& \hat U(t)  \ket{O(\beta)} \approx U_2(\vec{\theta}_2) \ket{O(\beta(\vec{\theta}))}.
\end{eqnarray}
After obtaining the above wavefunctions, the correlation function $C(t)$ can be obtained by computing the transition amplitude $\bra{\psi_2(t)} A \ket{\psi_1(t)}$. The specific quantum circuits for computing the transition amplitude can be found SM. 

\begin{figure}[ht!]
    \centering
  \includegraphics[width=0.50\textwidth]{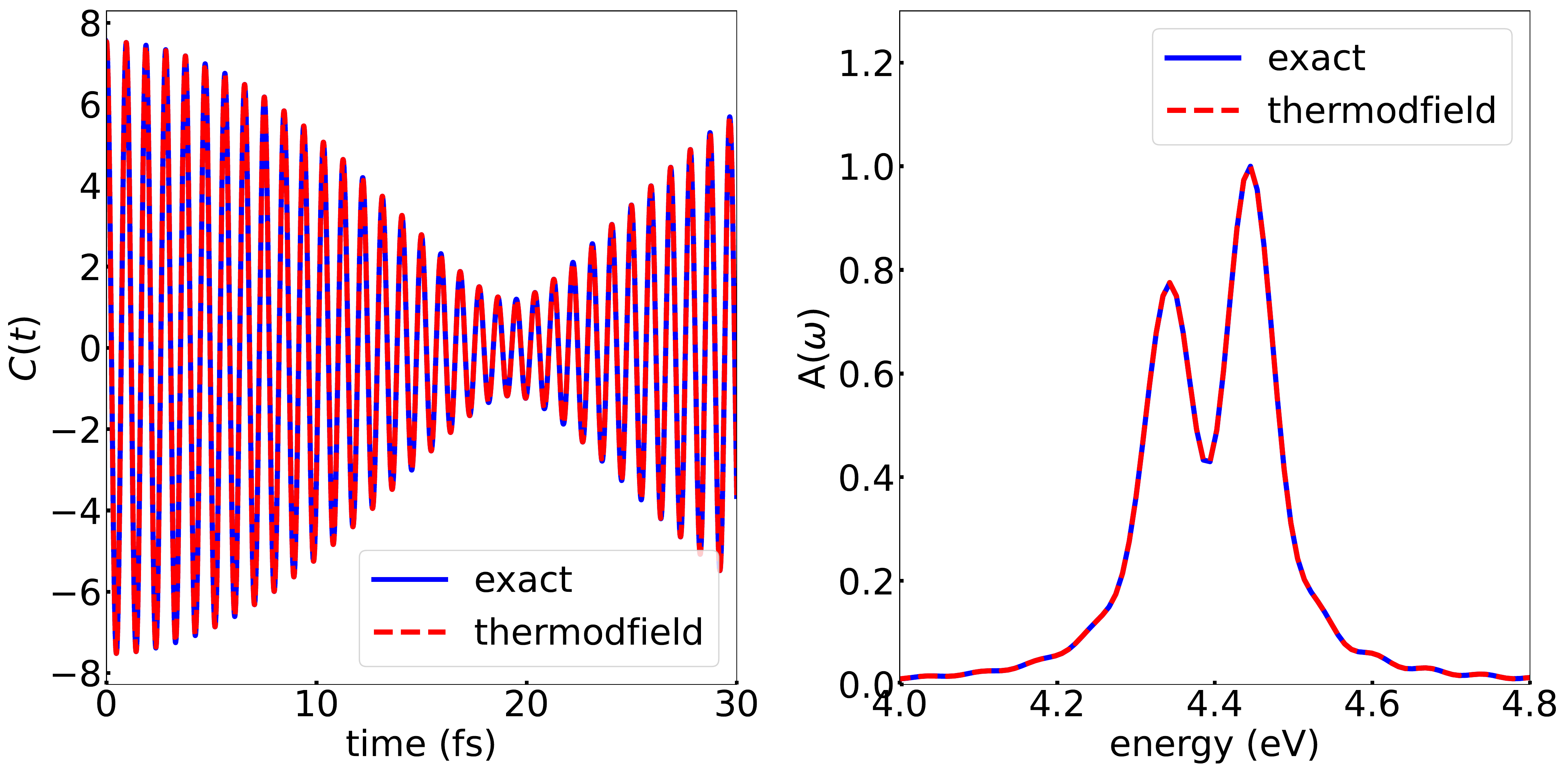}
    \caption{(left) Dipole-dipole time correlation function of a bi-thiophene tetramer at 300K. (right) UV-Vis absorption spectra obtained from the Fourier transform of the time correlation function. Results from the TFD-based variational quantum algorithm (blue dashed lines) are compared with numerically exact calculations (red solid lines).}
    \label{fig:spectrum}
\end{figure}

An important application of the finite-temperature correlation function is the calculation of optical absorption spectra of molecular multi-chromophoric systems, such as light-harvesting systems, as these spectra can provide a means to probe the underlying electronic processes in these systems.
In such spectroscopy calculations, the operators $A$ and $B$ are both the dipole operator and the absorption spectrum can then be obtained by taking the Fourier transform of the correlation function. 
As an example, we compute the UV-Vis absorption spectrum of a system consisting of interacting bi-thiophene molecules, 
in which each molecule is approximated as a two-level system. Bi-thiophene molecule represents the minimum model within the family of thiophene-based polymers, which have important applications in thin film technology, such as field-effect transistors. 
Here each bi-thiophene is represented by one qubit and the simulation details including the mapping of the molecular Hamiltonian to Pauli matrices are provided in the SM. The simulation results of the dipole-dipole correlation function and absorption spectrum at 300K are shown in Fig. \ref{fig:spectrum}. We again obtain good agreement with exact calculations, reaffirming the capability of the TFD-based variational algorithm. All the numerical simulations in this work were performed using the tensor-network-based quantum circuit simulator TensorCircuit~\cite{Zhang2022}.

\textit{Discussions} - Here we develop a quantum simulation protocol for computing dynamical quantum properties at finite temperatures.  
As compared to the static counterparts, we argue that dynamical finite-temperature phenomena are a more fitting regime to demonstrate the capability of quantum computers in simulating quantum many-body physics. There are a plethora of classical methods to compute the static equilibrium properties of quantum systems based on the quantum Monte Carlo methods. 
On the other hand, choices are more limited for equilibrium dynamical properties (e.g. time correlation functions) due to the presence of the evolution operator, $e^{-iHt}$, in the quantum dynamics. This leads to the alternating weight problem that severely limits the convergence of Monte Carlo calculations~\cite{Thirumalai1991, Krilov2001}. One possible option is the path integral Monte Carlo (PIMC) that numerically performs the analytic continuation of imaginary-time correlation functions to real time, but long and expensive PIMC simulation is needed in order to ensure the stability of the analytic continuation~\cite{Krilov2001}. Furthermore, classical algorithms for non-equilibrium finite-temperature phenomena, such as those based on the multiconfigurational time-dependent Hartree (MCTDH) approach, often come with significant computational cost.

In this work we focus on using the variational quantum algorithms for initial state preparation and time evolution as it is expected that such algorithms are more suitable for NISQ hardware. 
However, there could be cases in which it is difficult to choose wavefunction ansatz that is sufficiently expressive in representing the quantum state. Additionally, the notorious barren plateaus issue~\cite{McClean2018} can be an obstacle in the optimization step. 
Fortunately, the TFD framework offers the flexibility to circumvent the variational methods in the quantum simulation, albeit at a potential cost of deeper circuit depth. 
One non-variational alternative for preparing the initial thermal state, $\ket{O(\beta)}$, is the QITE algorithm that approximates imaginary time evolution with unitary operators over a domain of qubits and is ansatz independent. 
The time-evolution of the double states can be implemented using the standard Trotter algorithm or other non-variational NISQ algorithms for Hamiltonian simulations, such as Ref.~\cite{Lau2022}. 

It is worth noting that if we are only interested in the observables of the physical system, the density matrix $\rho$ and its evolution is invariant with respect to any unitary transformation applied to the fictitious space (see the SM for more discussion).\cite{paeckel2019} It is in fact possible to set $\tilde {H} = 0$ without affecting the expectation values of the observables of the physical system. However, it has been shown that choosing $\tilde {H}$ to be the same Hamiltonian as the physical system leads to significantly lower entanglement entropy during the time evolution.\cite{karrasch2012a,karrasch2013,hauschild2018a} Such reduction of entanglement has been shown to be advantageous in the implementation of the real-time variational simulation in the context of matrix product states, as the bond dimension needed is reduced~\cite{hauschild2018a}. 

To conclude, we have introduced a quantum simulation protocol to simulate the finite-temperature dynamical properties based on the TFD formalism in both the equilibrium and non-equilibrium regimes. Compared to the previous approach in computing the equilibrium dynamical properties, our approach does not require the expensive sampling of initial state. The sampling cost used in the QITE algorithm in principle scales exponentially with system sizes unless approximations are used. 
More importantly, our approach is capable of simulating non-equilibrium phenomena which have not been previously explored using quantum computers. 
Non-equilibrium finte-temperature physics is a topic of both fundamental and practical importance, e.g. in the studies of cold quantum gases~\cite{Proukakis2013}, and existing classical simulation methods are computationally expensive. 
Our work here thus opens the door for simulating these classically intractable problems with NISQ devices. 
Finally it is worth noting that while we focus on closed systems in this work, the TFD formalism could be extended to simulations of quantum open systems, and incorporating the open-system TFD formalism into quantum simulation will be a topic of future research. 
\bibliography{MyCollection,liang}
\end{document}


\title{Supplemental Material: Variational Quantum Simulations of Finite-Temperature Dynamical Properties via Thermofield Dynamics}

\author{Chee Kong Lee}
\email{cheekonglee@tencent.com}
\affiliation{Tencent America, Palo Alto, CA 94306, United States}
\author{Shi-Xin Zhang}
\affiliation{Tencent Quantum Lab, Shenzhen, Guangdong 518057, China}
\author{Chang Yu Hsieh}
\affiliation{Tencent Quantum Lab, Shenzhen, Guangdong 518057, China}
\author{Shengyu Zhang} 
\affiliation{Tencent Quantum Lab, Shenzhen, Guangdong 518057, China}
\author{Liang Shi}
\email{lshi4@ucmerced.edu}
\affiliation{Chemistry and Biochemistry, University of California, Merced, California 95343, United States}

\maketitle

\subsection{Implementation details of the variational quantum algorithms}
\begin{figure}[ht!]
    \centering
  \includegraphics[width=0.9\textwidth]{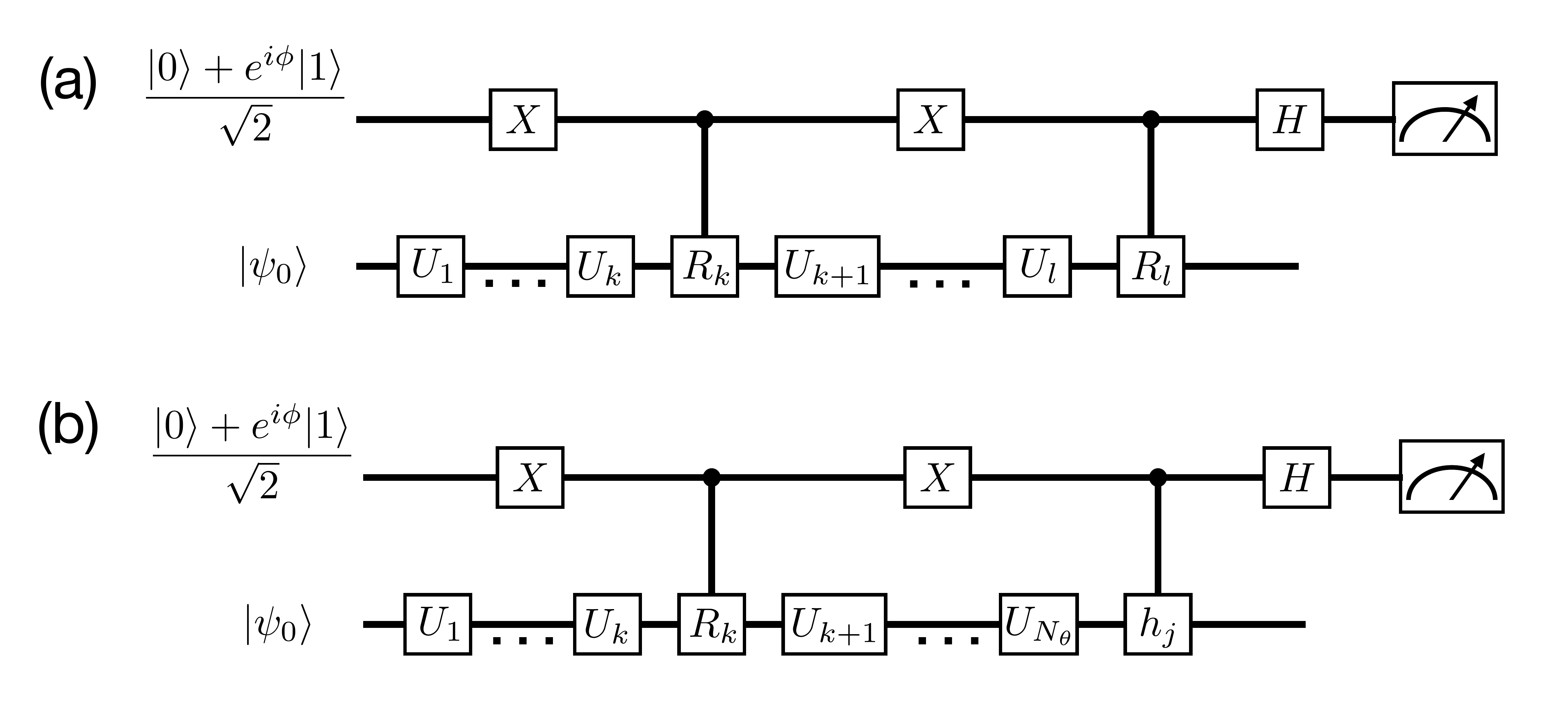}
    \caption{Quantum circuits to compute the matrix elements (a) $M_{kl}$ and (b) $ V_k$ in Eq.~(\ref{eqn:circuit}). The ancillary qubit is initialized in state $\frac{\ket{0} + e^{i\phi} \ket{1} }{\sqrt{2}}$. The phase factor $\phi$ is set to be $0$ or $\pi/2$ in order to measure the real and imaginary components, respectively.}
    \label{fig:circuit}
\end{figure}

In this section we show how the matrix elements of $M$ and $V$ in the main text can be measured in quantum circuits. 
We consider wavefunction ansatz of the form
\begin{eqnarray}
\ket{\psi(\vec{\theta})} = {U}(\vec \theta)\ket{\psi_0}  = \prod_{k=1}^{N_{\theta}} {U}_k(\theta_k) \ket{\psi_0}  =  \prod_{k=1}^{N_\theta} \mbox{e}^{i \theta_k {R}_k}  \ket{\psi_0}, 
\end{eqnarray}
where $\ket{\psi_0}$ is the initial state of the wavefunction. In the case of thermal state preparation with imaginary-time algorithm, $\ket{\psi_0} = \ket{I}$, 
${R}_k$ is some Pauli string, and $N_{\theta}$ denotes the number of variational parameters. 
The derivative of each term can be written as $\frac{\partial {U}_k(\theta_k)}{\partial \theta_k} = i {R}_k U_k(\theta_k)$, therefore the matrix elements of $M$ and $ V$ are (assuming $k<l$)
\begin{eqnarray} \label{eqn:circuit}
 M_{kl} &=&     \bra{\psi_0}  U_1^\dagger  ... U_k^\dagger  R_k^\dagger ... \hat U_L^\dagger
                         U_L ...  R_l  U_l ...  U_1  \ket{\psi_0}  \\
       &=&  \bra{\psi_0}  U_1^\dagger  ... U_k^\dagger  R_k^\dagger  U_{k+1}^\dagger  ... U_l^\dagger  R_l  U_l ...  U_1  \ket{\psi_0} ,  \nonumber \\
 V_{k}   &=&   i \sum_j c_j \bra{\psi_0}  U_1^\dagger  ... U_{N_\theta}^\dagger  h_j
                         U_{N_\theta} ...  R_k  U_k ...  U_1  \ket{\psi_0} ,   \nonumber         
\end{eqnarray}
where we have expressed the Hamiltonian as $ H = \sum_j c_j {h}_j$ where $h_j$ are some Pauli strings and $c_j$ are the corresponding coefficients. 
The real and imaginary parts of the matrix elements in Eq. (\ref{eqn:circuit}) can be obtained with quantum circuits via the Hadamard test, shown in Fig.~\ref{fig:circuit}.

In our simulations of the molecular and spin systems in the maim text, we include ${X, Z, XX, YY, ZZ}$ rotations, consider all-to-all couplings for the two-qubits rotations. 

\subsection{Quantum circuits for computing the time-correlation functions}
\begin{figure}[ht!]
    \centering
  \includegraphics[width=0.9\textwidth]{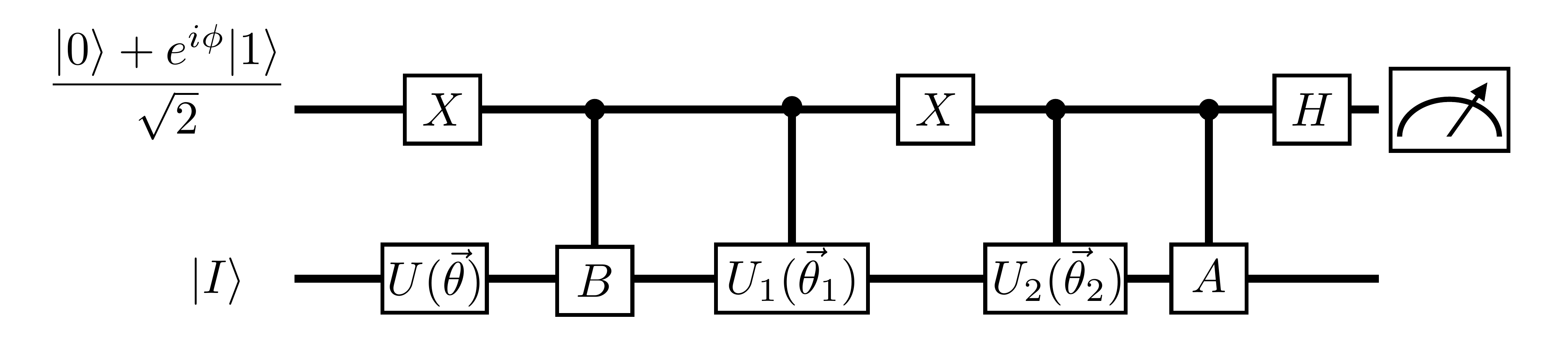}
    \caption{Quantum circuits to compute the time-correlation $C(t)$. The phase factor $\phi$ is set to be $0$ or $\pi/2$ in order to measure the real and imaginary components, respectively.}
    \label{fig:circuit_transition}
\end{figure}

Here we discuss how to evaluate the time-correlation function $C(t) = \bra{O(\beta)} A(t) B \ket{O(\beta)}$ in quantum circuits. As mentioned in the main text,
 we first prepare the thermal state using the imaginary-time VQA:  $\ket{O(\beta)} \approx \ket{O(\beta(\vec{\theta}))} = U(\vec{\theta}) \ket{I}$. We then propagate two sets of wavefunctions using the real-time VQA: 
\begin{eqnarray}
    \ket{\psi_1(t)} &=& U(t) B \ket{O(\beta)} \approx U_1(\vec{\theta}_1) B \ket{O(\beta(\vec{\theta}))}, \\
   \ket{\psi_2(t)} &=& U(t)  \ket{O(\beta)} \approx U_2(\vec{\theta}_2) \ket{O(\beta(\vec{\theta}))}. 
\end{eqnarray}
The correlation function $C(t)$ is simply the transition amplitude 
\begin{eqnarray}
    C(t) = \bra{\psi_2(t)} A \ket{\psi_1(t)}
    \approx \bra{I} U^\dagger(\vec \theta)  U_2^\dagger(\vec \theta_2)  A U_1(\vec \theta_1) B  U(\vec \theta)  \ket{I}. 
\end{eqnarray}
The right-hand-side of the above equation can be computed using the quantum circuit depicted in Fig.~\ref{fig:circuit_transition}~\cite{Endo2019}. 

\subsection{Overview of the thermal-field dynamics (TFD) theory}

In this section, we briefly review the TFD formulation of quantum mechanics. Consider a time-dependent Hamiltonian, $H(t)$, the density matrix, $\rho(t)$, satisfies the Liouville-von Neumann equation ($\hbar=1$):
\be
i \timeder \rho(t) = [H(t), \rho(t)].
\ee
Its formal solution is
\be
\rho(t) = U(t) \rho(0) U^\dagger(t),
\ee
where the time-ordered exponential is given by
\be
U(t) = \exp_+ \left [ -i \int_0^ t d\tau H(\tau) \right].
\ee
The square root of density matrix is then formally given by
\be
\rho(t)^{1/2} = U(t) \rho(0)^{1/2} U^\dagger(t),
\ee 
and its equation of motion is
\be
i \timeder \rho(t)^{1/2} = [H(t), \rho(t)^{1/2}].
\label{eq:eom_for_root_rho}
\ee

The TFD theory purifies a mixed state of the physical system by doubling the Hilbert space, where the fictitious system is identical to the physical one. The time-dependent thermal quantum state in the double space is defined as
\be
\ket{\Psi(t)} = d^{1/2} \rho(t)^{1/2} \ket{I},
\label{eq:def_tfd_state}
\ee
where $d$ is the dimension of the Hilbert space of the physical system, and 
\be
\ket{I} = \sum_n \ket{n}\ket{\tilde{n}}/\sqrt{d},
\ee
and $\ket{n}$ is the eigenstate of the physical system Hamiltonian ($H$) with $\ket{\tilde{n}}$ being the corresponding eigenstate of the fictitious one ($\tilde{H}$). We use the tilde notation for the fictitious system. With a unitary transformation, $\ket{I}$ can also be expressed in any complete orthonormal basis, $\{\ket{\alpha}\}$,
\be
\ket{I} = \sum_\alpha \ket{\alpha}\ket{\tilde{\alpha}}/\sqrt{d}.
\ee
As the Hamiltonians of the physical and fictitious systems are identical and commute, it is straightforward to derive the equation of motion for $\ket{\Psi(t)}$ from Eq. (\ref{eq:eom_for_root_rho}) and (\ref{eq:def_tfd_state}):\cite{suzuki1985}
\be
i \timeder \ket{\Psi(t)} =  \hat{H}(t) \ket{\Psi(t)},
\label{eq:eom_tfd}
\ee
where $\hat{H}(t) = H(t) - \tilde{H}(t)$. For an arbitrary operator, $A$, in the physical system, its statistical average with $\rho(t)$ becomes the corresponding expectation value with $\ket{\Psi(t)}$,
\be
\avg{A}_t = \textrm{Tr}\{\rho(t) A\} = \bra{\Psi(t)} A \ket{\Psi(t)},
\label{eq:avg_t}
\ee
where the subscript $t$ indicates the time dependence of the statistical average. The density operator can be recovered by taking the trace over the fictitious system, $\rho(t) = \textrm{Tr}_f \ket{\Psi(t)}\bra{\Psi(t)}$.

If the physical system is at thermal equilibrium, $\rho(t) = \rho_{\textrm{eq}} = e^{-\beta H} / Z$, where $Z = \textrm{Tr} \{e^{-\beta H} \}$, and $\beta = 1/k_B T$. $\ket{\Psi(t)}$ now becomes the thermo-field double state (also known as thermal vacuum state), 
\be
\ket{O(\beta)} = d^{1/2} Z(\beta)^{-1/2} e^{-\beta H/2} \ket{I}.
\ee
The thermal equilibrium average of an arbitrary operator, $A$, of the physical system is then given by
\be
\avg{A} = \textrm{Tr} \{ \rho_{\textrm{eq}} A \} = \bra{O(\beta)}  A  \ket{O(\beta)}.
\ee

It is worth pointing out that the density matrix, $\rho(t)$, and Eq. (\ref{eq:avg_t}) are invariant with respect to arbitrary unitary transformation, $\tilde{V}$, applied to the fictitious space. The equation of motion for the transformed thermal quantum state, $\ket{\Psi_V(t)} \equiv \tilde{V} \ket{\Psi(t)}$, keeps the same form as Eq. (\ref{eq:eom_tfd}) but with a re-defined $\hat{H}$, namely $\hat{H}_V \equiv H - \tilde{V} \tilde{H} \tilde{V}^\dagger$. However, since the evolution of the fictitious system does not affect the density matrix of the physical system, one can still evolve the thermal quantum state using $\hat{H} = H - \tilde{H}$, which is sometimes a helpful choice.\cite{karrasch2012a,karrasch2013,hauschild2018a}

\subsection{Simulation details for the absorption spectrum of bi-thiophene (T2) tetramer}

In this section, we provide the details for the UV-Vis absorption spectral simulation of a model T2 tetramer. Each T2 molecule can be considered as a two-level system with the ground state and the first excited state, denoted as $\ket{0}$ and $\ket{1}$, respectively. The electronic Hamiltonian of the tetramer, which involves one-body and two-body interactions, can be expressed in the basis of the product states,  $\ket{p_1}\otimes\ket{p_2}\otimes\ket{p_3}\otimes\ket{p_4}$, where $p_i$ is 0 or 1. If each T2 is represented by a qubit, following the work by Parrish and co-workers,\cite{parrish2019} the electronic Hamiltonian can be written in terms of Pauli matrices:
\begin{eqnarray} 
\hat{H}(t) = \mathcal{E} \hat I + \sum_{m=1}^4 [ \mathcal{Z}_m \hat Z_m + \mathcal{X}_m \hat X_m ] + \sum_{m=1}^4 \sum_{n<m} [\mathcal{XX}_{mn} \hat X_m \otimes \hat X_n + \\
\mathcal{XZ}_{mn} \hat X_m \otimes \hat Z_n + \mathcal{ZX}_{mn} \hat Z_m \otimes \hat X_n + \mathcal{ZZ}_{mn} \hat Z_m \otimes \hat Z_n ], \nonumber
\label{eq:H_martinez}
\end{eqnarray}
where $\hat I$ is the identity operator, $\hat X_m$, $\hat Y_m$, and $\hat Z_m$ are the Pauli operators associated with the $m$th T2. The coefficients for the corresponding Pauli operators in the Hamiltonian, $\{\mathcal{E},\mathcal{Z}_m,\mathcal{X}_m,\mathcal{XX}_{mn}, \mathcal{XZ}_{mn}, \mathcal{ZX}_{mn}, \mathcal{ZZ}_{mn}\}$, are given by
\begin{eqnarray} 
\mathcal{E} = \sum_m S_m + \sum_{n<m} (S_m|S_n)\\
\mathcal{Z}_m = D_m + \sum_n (D_m|S_n) \\
\mathcal{X}_m = X_m + \sum_n (T_m|S_n) \\
\mathcal{XX}_{mn} = (T_m|T_n) \\
\mathcal{XZ}_{mn} = (T_m|D_n)\\
\mathcal{ZX}_{mn} = (D_m|T_n)\\ 
\mathcal{ZZ}_{mn} = (D_m|D_n).
\end{eqnarray}
In these expressions, $S_m$, $D_m$, and $X_m$ are determined by the matrix elements of one-body term, $\hat h$, associated with the $m$th T2,
\begin{eqnarray} 
S_m = [ ( 0_m | \hat h | 0_m ) + ( 1_m |  \hat h | 1_m ) ] /2 \\ 
D_m = [ ( 0_m | \hat h | 0_m ) - ( 1_m |  \hat h | 1_m ) ] /2\\
X_m = ( 0_m | \hat h | 1_m ),
\end{eqnarray}
where $| 0_m )$ and $| 1_m )$ are the ground and excited states of the $m$th T2. The terms of the form $(\cdots|\cdots)$ are associated with the two-body interaction term, $\hat v$, and symbols $|S_m)$, $|D_m)$, and $|T_m)$ denote
\begin{eqnarray} 
|S_m) = |0_m 0_m + 1_m 1_m )/2 \\ 
|D_m) = |0_m 0_m - 1_m 1_m )/2\\
|T_m) = |0_m 1_m).
\end{eqnarray}
For example, $(T_m | S_n) = (0_m 1_m | \hat v | 0_n 0_n + 1_n 1_n)/2$. 

To construct this electronic Hamiltonian, we performed molecular dynamics (MD) simulation of a T2 crystal consisting of 64 T2 molecules in the simulation box, and took four T2 molecules from a random configuration in the MD trajectory. The setup of the MD simulation was described in detail in our previous work\cite{Lee2022}.
We then computed the lowest-lying excited-state energy, ground-state dipole, excited-state dipole, and transition dipole for each T2 molecule using density functional theory (DFT) and time-dependent DFT (TDDFT) with the Tamm-Dancoff approximation at the theory level of CAM-B3LYP/6-31+G(d). All the quantum chemical calculations were performed with the PySCF program,\cite{sun2018} and density fitting was used with the heavy-aug-cc-pvdz-jkfit auxiliary basis set implemented in PySCF. For the coefficients of the one-body terms in Eq. (\ref{eq:H_martinez}), we approximately set $( 0_m | \hat h | 0_m )=( 0_m | \hat h | 1_m )=0$, and $( 1_m | \hat h | 1_m )$ to be the lowest lying excited state energy of the $m$th T2 molecule. The coefficients for the two-body terms were computed using the approximation of dipole-dipole interaction, and the dipole position was taken to be the center of mass of the T2 molecule. Ground-state, excited-state and transition dipoles were used for the terms involving $|0_m 0_m)$, $|1_m 1_m)$, and $|0_m 1_m)$, respectively. For example, 
\be
(0_m 0_m| \hat v | 0_n 1_n) \approx \frac{\vec{\mu}_{00,m} \cdot \vec{\mu}_{01,n} - 3(\vec{\mu}_{00,m} \cdot \hat{r}_{mn})(\vec{\mu}_{01,n} \cdot \hat{r}_{mn}) }{r_{mn}^3}, 
\ee
where $\vec{\mu}_{00,m}$ is the ground-state dipole of the $m$th T2, $\vec{\mu}_{01,n}$ is the transition dipole associated with the excitation on the $n$th T2, $\vec{r}_{mn}$ is the vector connecting the centers of mass of the two T2 molecules, $r_{mn} = |\vec{r}_{mn}|$, and $\hat{r}_{mn} = \vec{r}_{mn} / |\vec{r}_{mn}|$. By constructing the electronic Hamiltonian from atomistic simulations, we hope to include some realistic disorder in the Hamiltonian.

To compute absorption spectrum, we also need the electric dipole operator, which can be constructed in a similar manner, and the $k$th Cartesian component ($k=x,y,z$) of the dipole operator is given by\cite{parrish2019}
\be
\hat \mu^k(t) =  \sum_m [  \mu_{I,m}^k \hat I_m + \mu_{Z,m}^k \hat Z_m + \mu_{X,m}^k \hat X_m ] ,
\label{eq:mu_martinez}
\ee
where the cofficients are 
\begin{eqnarray} 
\mu_{I,m}^k = (\mu_{00,m}^k + \mu_{11,m}^k)/2 \\ 
\mu_{Z,m}^k = (\mu_{00,m}^k - \mu_{11,m}^k)/2\\
\mu_{X,m}^k = \mu_{01,m}^k,
\end{eqnarray}
where $\mu_{00,m}^k$, $\mu_{11,m}^k$, and $\mu_{01,m}^k$ are the $k$th Cartesian components of the ground-state, excited-state and transition dipoles for the $m$th T2.

For simplicity, we assume that the incident light is polarized in the $x$ direction such that we only need to consider the time correlation function of the $x$ component of the dipole operator, namely $A=B=\mu_x$ in Eq. (14) of the main text. To reduce spurious effects in the Fourier transform,\cite{Valleau2012,loco2019} we applied an exponential damping function to the time correlation function, $C(t)$, before performing the Fourier transform, and the absorption line shape is given by
\be
I(\w) \sim \int_{-\infty}^{\infty} dt\; e^{i\w t} C(t) e^{-|t|/\tau}, 
\label{eq:ft_lt}
\ee
where $\tau$ is the decay constant. In some cases, this damping function can also be used to phenomenologically include lifetime effects. In this work, we chose $\tau = 20$ fs.

Finally, it is worth pointing out that this approach of computing spectra of multi-chromophoric systems takes into account all possible excitations in the Hilbert space spanned by the products states, which in principle leads to an exponential scaling of the classical computational cost with the system size (i.e., number of T2 molecules in our case). This treatment is in contrast to the popular Frenkel exciton model, where electronic excitation is restricted to single-excitation manifold. The full Hilbert space approach becomes necessary when state-mixing or multiexciton effects are substantial,\cite{Levine2006,parrish2019,Spano1991,Renger1997,Bruggemann2001,May2014} and quantum computing may be advantageous. 



\bibliography{MyCollection,liang}